\documentclass[vecphys]{svmult}

\usepackage{makeidx}         % allows index generation
\usepackage{graphicx}        % standard LaTeX graphics tool
\usepackage{multicol}        % used for the two-column index
\usepackage[bottom]{footmisc}% places footnotes at page bottom

\makeindex             % used for the subject index
\begin{document}

\title*{Pulsars as Fantastic Objects and Probes} \author{JinLin Han}
\institute{National Astronomical Observatories, Chinese Academy of Sciences\\
  Jia-20 DaTun Road, Chaoyang District, Beijing 100012, CHINA \\
  \texttt{hjl@bao.ac.cn}}

\maketitle
 
\begin{abstract}
Pulsars are fantastic objects, which show the extreme states of
matters and plasma physics not understood yet. Pulsars can be
used as probes for the detection of interstellar medium and even the
gravitational waves. Here I review the basic facts of pulsars which should
attract students to choose pulsar studies as their future projects.
\end{abstract}

\section{Pulsars: General Introduction}

Pulsars are sending us pulses -- we can receive these pulses in radio 
bands. A small number of pulsars also emit high energy radiation in 
optical and X-ray or even $\gamma$-ray bands. Here let me discuss the 
radio pulsars, and leave the high energy emission and its explanation 
to Prof. Qiao in this volume.

After the first pulsar discovered by Hewish et al.  \cite{hbp+68} in
1968, it was soon realized that they are rotating neutron stars
\cite{gold68,pac68} with a diameter of only 20~km but extremely high
density ($10^{15}$g~cm$^{-3}$) and extremely strong magnetic fields
($10^8$ to $10^{14}$~G). Because of revealing of this new state of
matter in the universe, the pulsar discovery was awarded the Nobel
Prize in 1974 in physics.

Pulsars take birth in supernova explosion, which is evident from the
young pulsars and supernova remnants associations \cite{kas98}. For
example, the Vela pulsar is located in the center of Vela nebula, the
Crab pulsar is acting as the heart of Crab nebula. Pulsars get high
velocity (a few 100~km~s$^{-1}$) \cite{hllk05,wlh06} in the explosion
so that pulsars are running away quickly from their birthplace, even
about half pulsars have escaped from our Galaxy in last 100~Myr
\cite{sh04}.

The broadband radio emission of pulsars leaves from the emission
region almost simultaneously. However, after these signals pass
through the interstellar medium (ISM), the radio waves at a lower
frequency, $\nu_1$, in GHz, come later than these at a higher
frequency $\nu_2$, with a time delay of
\begin{equation}
dt = 4.15\left ({1\over \nu_1^2} - {1\over \nu_2^2}\right) DM~~~{\rm ms},
\end{equation}
where $DM$ is known as the dispersion effect from the ionized gas in the ISM
along the path from a pulsar to us, given by 
\begin{equation}
DM=\int_0^D n_e dl~~~\mbox{pc~cm$^{-3}$}, 
\end{equation}
where $n_e$ is the electron density in units of cm$^{-3}$ and $D $ the
distance from observer to pulsar in pc.  In Fig.~1 (left panel), we
have plotted the signals received in each frequency channel as
function of pulse phase. It is evident that the signals at low
frequency are delayed in arrival time compared to those at high frequencies.
After making the delay corrections and adding the channels we get the
pulse profile (see Fig.~1 right panel).
\begin{figure}
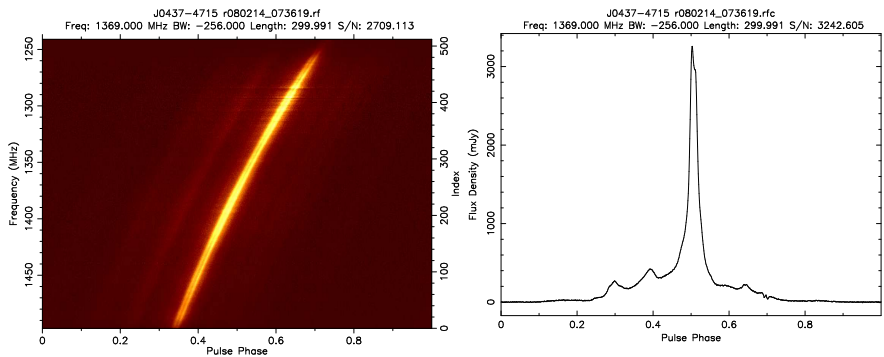

\centering
\includegraphics[height=60mm,angle=270]{Han_f1a.ps} 
\includegraphics[height=55mm,angle=270]{Han_f1b.ps} \\
\caption{The dispersion of pulsed signals -- the example is observation 
of PSR J0437-4715 (left). The dedispersion and adding signals from all frequency 
channels gives a strong and stable profile (right). Data is obtained by the
author with Parkes telescope.}
\label{fig:1}       % Give a unique label
\end{figure}

How to find pulsars? Because a pulsar have a very accurate period -- the
period change due to slow-down is small enough in half an hour, one can make
Fourier transform of the recorded power, and found the period in the
power-spectrum. Nowadays, there are many terrestrial radio frequency
interferences (RFI) which look like pulsar signals. However, these RFI
normally show the maximum power at zero dispersion. Therefore, when radio
power from many frequency channels is recorded, only after the power from
all channels are properly de-dispersed the pulsar signal should show the
maximum power. Therefore, here are several steps to find a new pulsar: {\bf
1)} record signals in time series at many frequency-channels; {\bf 2)}
de-disperse the channel signals with many trial DM and add all channels
together to get one time series for each trial DM; {\bf 3)} search for the
periodicity from each dedispersed time series. Because most pulsars have
only narrow pulses, the power-spectrum should show not only the primary
pulsar rotation frequency, but also its harmonics. Often the harmonics are
added together to enhance the searching signal-to-noise ratio; {\bf 4)}
verify the dedispersed pulse by searching for the best detection
signal-to-noise ratio in the 2-D parameter space around the proposed $P$ and
$DM$ and then folding data in the right period $P$ and $DM$; {\bf 5)}
finally re-observe again in this sky position and search the pulse again
around the proposed $P$ and $DM$. If the pulse can be found in the same DM
but slightly evolved period, then a pulsar is definitely found! See
 \cite{mlc+01,cfl+06} for the pulsar searching strategy.

Up to now, about 1800 pulsars have been discovered  \cite{mht+05}, including
some discovered using the GMRT  \cite{fgr+04,jmk+07}. Most of pulsars are in
our Milky way Galaxy, and only about 20 were discovered from extensive
searches of nearby galaxies, the Large and Small Magellanic clouds. Pulsars
have a period of 1.3~ms to $\sim$10~s. Some very fast rotating pulsars,
so-called millisecond pulsars, have a period of only some milliseconds but
very stable and very small period derivatives. From measurements of binary
pulsars, it has been established that neutron stars normally have a mass
 \cite{pr06} about 1.4 solar mass ($M_{\odot}$) though some are heavier
(probably up to 2 $M_{\odot}$) and some are lighter (1.2 $M_{\odot}$).

Here are pulsar books I would like to recommend to readers: 1). ``Handbook
of Pulsar Astronomy'', by D. Lorimer and M. Kramer \cite{lk05}, published by
the Cambridge University Press (2005), contains many up-to-dated material of
pulsar studies; 2) ``Pulsar Astronomy'' by A. Lyne \& F. Graham-Smith
 \cite{lg06}, also published by the Cambridge University Press (2006), is an
excellent text book for graduate students, covering all aspects of pulsar 
astrophysics.

\begin{figure}
\centering
\includegraphics[width=116mm]{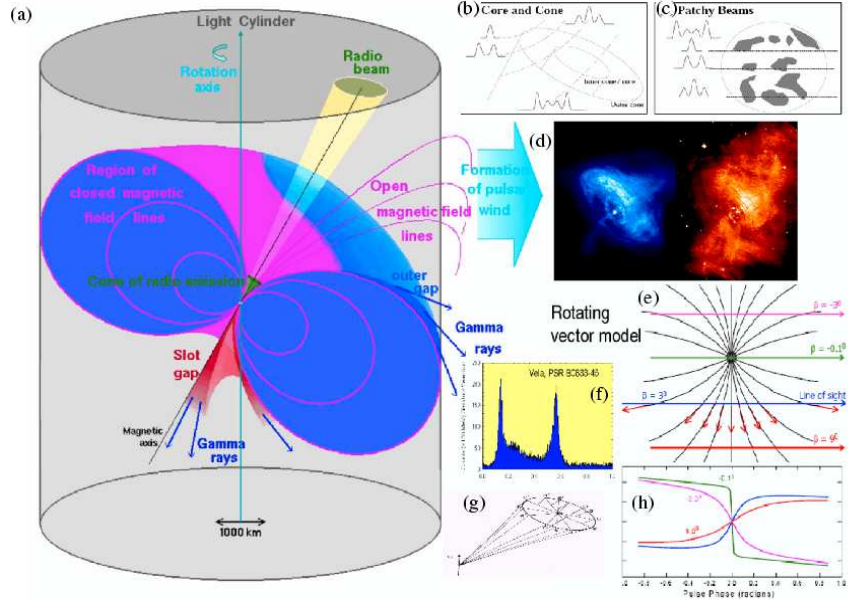} 
\caption{The magnetosphere and radiation of neutron stars: (a) the particles are
accelerated in the inner gap, outer gap or the slot gap, and radio emission
beams out near the magnetic poles. The radio beam is not uniformly illuminated
but maybe more bright in (b) conal regions or (c) randomly in patches. The
outflowing particles can form the pulsar wind nebula. The polarization 
of the radio emission depends on (e) the magnetic field planes -- when the
(g) line of sight impacts the beam, the observed polarization angles show
variations in the (h) ``S-shaped'' curves. This demonstration is composed by using 
the artwork produced by Drs. B. Link, D. Lorimer, A. Harding, G.J. Qiao, 
as well as (d) the Chandra and HST images of the Crab pulsar nebula,  and (f) the
$\gamma$-ray profile of the Vela pulsar.
}
\label{fig2}       % Give a unique label
\end{figure}

\section{Pulsar emission} 

Radio emission from pulsars is generated in pulsar magnetosphere. We define
the boundary of this magnetosphere by the light-cylinder, e.g. at the radius
where the rotation speed is equal to the light speed. The particles,
i.e. positrons and electrons, are accelerated along the magnetic fields
above polar cap or the outer gap. These particles radiate \cite{gan04} so
that we can see the emission in radio and high energy band. {\it However, it
is not clear what physical processes are involved for the particles to
radiate}. Pulsar wind or wind nebula \cite{sla08} can be formed if particles
flow out through the open magnetic field lines passing through the
light-cylinder.

It is the rotation that provides the energy source for pulsar emission and
particle outflowing. One can calculate the braking torque due to magnetic
dipole radiation, which finally result in $\dot{\mu}=-K \mu^n$, here the
$\mu$ is the rotation frequency of the neutron star, and the $n$ is the
braking index which theoretically should be equal to 3. However, the
measurements show that it is usually significantly smaller than 3, which
implies  \cite{yxz07} that other reasons are also consuming the rotation
energy, e.g. pulsar wind. Otherwise, magnetic fields may evolve. The
smaller index means that the magnetic fields may be growing \cite{lz04}.
In Fig.~\ref{fig2}, I provide an over view on pulsars, and models in
understanding their action.

\subsection{Pulse profiles and emission beam} 

Pulsars emit broadband radio waves. We receive these signals when the
emission beam is sweeping towards us. The line of sight impacts the emission
beam so that we see a pulse profile. Now it is clear that {\bf 1)} the
average pulse profiles are very stable, except for a few pulsars with
precession \cite{sls00,wt02}; {\bf 2)} profiles are the finger-prints of
pulsars which differ from each other. Most pulsar profiles have one or two
or three distinctive components or peaks, a small number of pulsars have 4
or 5, and occasionally more than 10 components \cite{mh04}. Some components
are difficult to be identified, but can be revealed by multi-Gaussian
fitting \cite{wgr+98} or ``window-threshold technique'' \cite{gg01}; {\bf
3)} these profiles vary with frequency, and each component has some-how
independent spectral index \cite{mh04}, although in general pulsars have
very steep spectrum \cite{mkkw00}; {\bf 4)} some pulsars have a main pulse
and an interpulse, separating about $180^{\circ}$ in the rotation phase,
which look like the emission from the two opposite magnetic poles; {\bf 5)}
Some pulsars show profiles in two or three modes, which is so-called
``mode-changing''. Some pulsars even switch off their emission for sometime,
which is called ``nulling'' \cite{wmj07}.

\begin{figure}[bt]
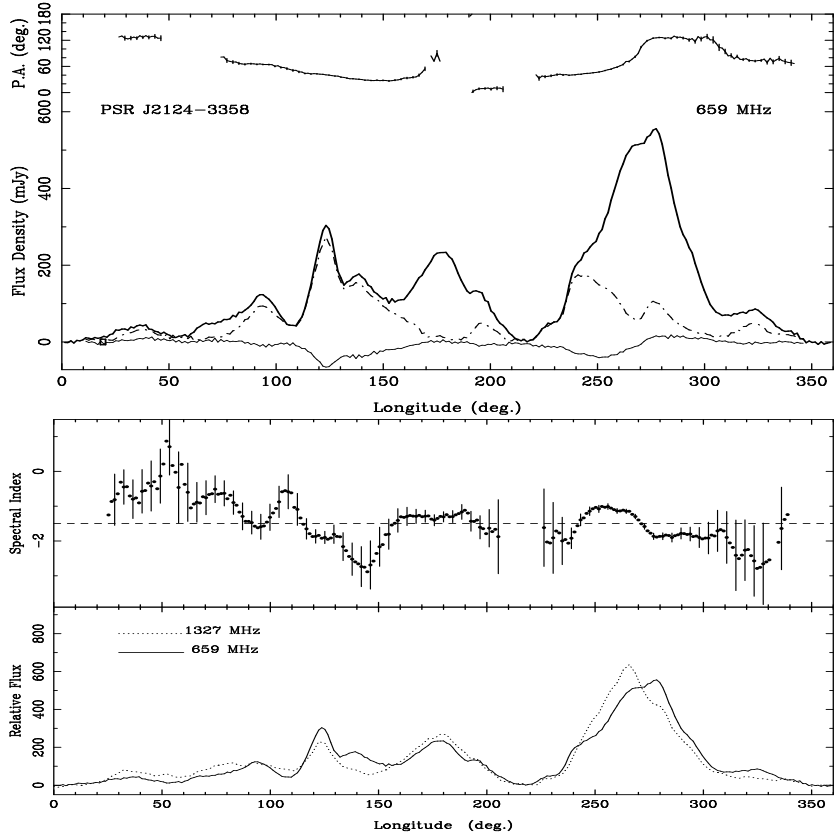

\centering
\includegraphics[height=11cm,width=5.5cm,angle=270]{Han_f3a.eps} \\
\includegraphics[height=11cm,width=5.5cm,angle=270]{Han_f3b.eps} \\
%\picplace{5cm}{2cm} % Give the correct figure height and width in cm
\caption{The polarization profile of PSR J2124-3358 at 659~MHz (top
panel) shows more than 12 distinguishable components, each with different
fraction of linear (dash line) and circular polarization (lower solid
line). Comparison of profiles at 659~MHz and 1327~MHz (lower panel) show
that the spectrum varies against rotation phase. This is an extraordinary 
profile (data from \cite{mh04}) and usually pulsars are not
so complicated -- see Fig.\ref{fig4}.}
\label{fig3}       % Give a unique label
\end{figure}

It is naturally understandable that the profile peaks indicate the bright
parts of the emission beam. That is to say, the pulsar emission beam is not
uniformly illuminated, some parts brighter, some fainter. The line of sight 
only impacts one slice of beam for a given pulsar, so it is not
possible to know the whole beam of any pulsar, except that one can fly in
space (!) and observe a pulsar in different lines of sight with respect to
its rotational axis! However, based on profiles of many pulsars, the
emission beam has been suggested ideally to consist of one core and nested
cones  \cite{ran83,gg01}. This image gained some good support from the
observational fact of the widening of the double profiles at lower
frequencies \cite{tho91}. This is so-called ``radius-frequency mapping'', which
is explained as the emission comes from a pair cuts of the ``outer
cone'' formed in the open dipole magnetic fields. However, if all peak
emission comes from cones, how many cones are needed to explain more than 10
components of some pulsars? Can these emission cones be really formed above
the magnetic polar caps? This problem can be eased in the so-called
``patchy beam'' model  \cite{lm88}, where the emission comes from many
bright parts of a beam. If the slices of emission beams of all pulsars
are put together, one can not see distinct cones  \cite{hm01}. 

In fact, the averaged pulse profiles can only be used to reveal emission
geometry or brightness distribution of the emission beam. A large amount of
information on the emission process can be obtained from observations of
individual pulses  \cite{dr99,dr01}. 
It has been established that for pulsars with drifting subpulses,
some emission zones is stably circulating around the magnetic axis of some
pulsars with remarkably organized configuration. A detailed modeling hints
that the imagined emission cones are patchy! To my understanding, {\it ``the
patchy cones''} seems to be the best description of the characteristics of
pulsar emission beam: the cones only roughly define the emission region in
the magnetosphere and ``patches'' are related to the non-random but preferred
bunches of field lines for generation of emission spots which are probably
physically related to sparking near the polar cap. This idea is similar but
different from the idea of ``window+sources'' proposed by Manchester in 1995
 \cite{man95}.

\subsection{Polarization}

\begin{figure}[bt]
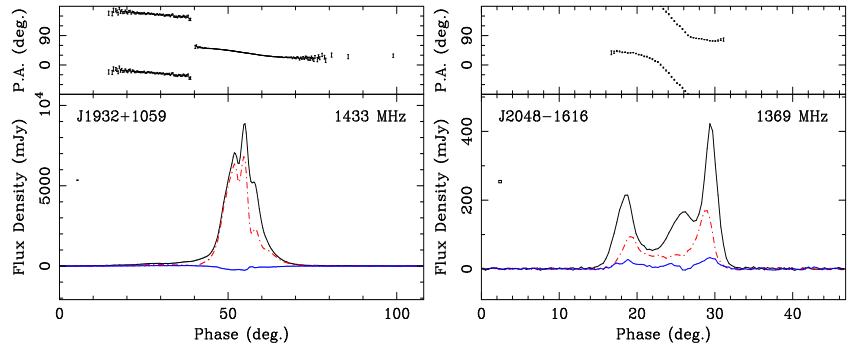

\centering
\includegraphics[height=55mm,width=45mm,angle=270]{Han_f4a.ps}
\includegraphics[height=55mm,width=45mm,angle=270]{Han_f4b.ps}
%\picplace{5cm}{2cm} % Give the correct figure height and width in cm
\caption{The polarization profile of PSR J1932+1059(=B1929+10) main pulse
and J2048$-$1616 at 1369~MHz, observed by the author using Parkes telescope.
The dashed line is the linear polarization intensity, and the circular
polarization is plotted by the lower solid line. The polarization angle
curve is plotted on the top of a profile.
}
\label{fig4}       % Give a unique label
\end{figure}

Pulsars are the strongest polarized radio sources in the universe. Almost
100\% linear polarization is detected (see Fig.~\ref{fig4}) for the whole or
a part of profiles of some pulsars \cite{wmlq93,mhq98,wcl+99}. In fact, it
is the polarization angle sweeping of the Vela pulsar that leads to the
famous ``rotating vector model'' proposed by Radhakrishnan and Cooke in 1969
\cite{rc69}, which solidly established that the radio emission comes from
region not far away from the magnetic poles of the neutron stars, and the
observed polarization angle is related (either parallel or perpendicular) to
the plane of magnetic field lines, which should have a ``S''-shape (see
PSR J2048$-$1616).

Assuming magnetic fields of pulsars are dominated by the dipole fields,
then one can determine the emission geometry from the polarization
observations \cite{ran83}. The maximum polarization angle swept rate gives
the information on the smallest impact angle of the line-of-sight from the
magnetic axis \cite{lm88}. After classification of pulsar profiles, Rankin
has made extensive efforts to pin down the emission regions, by calculating
the geometrical parameters for various types of pulsars  \cite{ran83,ran90,
ran93a,ran93b}. The geometrical parameters of a large number of pulsars were
also calculated by Gould \& Lyne  \cite{gl98}.  From the phase shift between
the pulse center and the largest sweep rate of polarization angle curve, the
emission latitude can be estimated  \cite{bcw91,gan05}.
 
The polarization angle curves often are not smooth but have some jumps (see
PSR J1932+1059 in Fig.~\ref{fig4}).  When the polarization data samples of
individual pulses are plotted against the phase bins, the orthogonal
polarization modes can be solidly identified
\cite{man75,crb78,scr+84,scm+84}. Several possible origins of the orthogonal
polarization modes have been suggested. These orthogonally polarized modes
may reflect the eigenmodes of the magneto-active plasma in the open magnetic
field lines above the pulsar polar cap, with different refraction indexes
\cite{ba86} to separate these modes in outwards emission, or with different
conversion of different modes \cite{pet01}. It is also possible that the
pulsar emission of two modes is generated in one emission region by
positrons and electrons respectively \cite{am82,gan97}, i.e. the intrinsic
origin from the emission process rather than the propagation process. The
non-orthogonal emission modes also have been observed from some pulsars
\cite{glr+92}, which could be the superposition of emission of two modes
\cite{mmkl06} or two regions \cite{xqh97}.

Very special to the pulsars is the circular polarized emission, unique in
the universe. Usually it is as strong as 10\%, but in some pulsar components
it could reach 70\%  \cite{mh04}. The circular polarization measurements have
been comprehensively reviewed in  \cite{hmxq98,yh06}. In summary, we found
that circular polarization is not restricted to core components and, in some
cases, reversals of circular polarization sense are observed across the
conal emission. For core components, there is no significant correlation
between the sense change of circular polarization and the sense of linear
position-angle variation. These results are contradictory to the conclusions
given by Radhakrishnan \& Rankin  \cite{rr90} based on early smaller sample
of pulsar data. We found that in conal double profiles, the sense of
circular polarization is correlated with the sense of position-angle
variation. Pulsars with a high degree of linear polarization often have one
hand of circular polarization across the whole profile. For most pulsars,
the sign of circular polarization is same at 50~cm and 20~cm
wavelengths, and the degree of polarization is similar, albeit with a wide
scatter. Some pulsars are known to have frequency-dependent sign reversals.
The diverse behavior of circular polarization may be generated in the
emission process  \cite{ggm01} or arise as a propagation effect  \cite{ml04}.

\section{Pulsars as probes for interstellar medium}

The dispersion of the radio pulses is a very important tool, which can be 
used to not only identify radio emission from distant pulsars but also probe
the ionized gas in the interstellar medium (ISM) which otherwise is
difficult to detect and model. The scattering of radio signals 
measured from narrow radio pulses can be used to detect the
randomness of electron distribution. The polarized radio signals of pulsars
are Faraday rotated in the magnetized interstellar medium, so that pulsars
act as key probes for the Galactic magnetic fields.

\subsection{Electron density distribution in our Galaxy}

The most knowledge of interstellar electron distribution comes from pulsar
measurements. Using the observed dispersion measures of a sample of pulsars,
$DM=\int_{\rm 0}^{\rm Dist} n_e dl$, one can get the electron density
distribution if pulsar distances can be independently measured; or vice
versa. The models for electron density distribution of our Galaxy have been
constructed and constrained from the DMs and independent distance estimates
of a large number of pulsars \cite{tc93,gbc01,cl02,cl03}, which consists of
the thin electron disk, thick disk and spiral arms as well as bulge
component.  It is such an electron density model that can be used to
estimate distances of most of pulsars from observed DMs so that we can know
approximately how far a pulsar is located.

In fact, the interstellar medium is not smoothly distributed, but more or
less random or irregularly clumpy, with only some large-scale preferred
distribution in spiral arms and concentrated towards the Galactic plane.
Note also that pulsars are point radio sources. They move very fast
 \cite{hllk05}. The interstellar medium randomly refracts and diffracts the
pulsar signal and makes the pulsar scintillating and scattering. When
pulsars are observed in many frequency channels for some time, one can see
that pulsar signals scintillate both in time and observation channel
frequency dimension -- see excellent examples of so-called dynamic spectrum
shown by Gupta et al.  \cite{grl94}.  The dynamic spectrum, which itself is
the grey-plot of pulse intensity in the 2-D of observation frequency and
time, shows the intensity correlation over both the time scale and frequency
scale. Note that the randomness of the interstellar medium can be averaged
over certain distance, so the distant pulsars do not scintillate much. The
scintillation bandwidth is a function of frequency  \cite{wmj+05}. High
sensitivity observations can produce parabolic arcs in the secondary
spectrum of the dynamic spectrum  \cite{smc+01}, which depends on the
velocity of scintillation pattern and geometry of the diffracting screen.

The more distant the pulsar is located,  more the pulse signal is
diffracted and scattered. This leads to the pulse broadening, especially at
low frequencies. The most recent and largest data set for pulsar scattering
have been obtained by Bhat et al  \cite{bcc+04}, who observed 98 pulsars in
several frequencies, from which, together with previous the pulse-broadening,
$\tau$, is found to scale with observation frequency $\nu$ and DM, in the
form of
\begin{equation}
\log \tau {\rm (ms)} \simeq -6.46 + 0.154 \log(DM) + 1.07 \log(DM)^2 -
4.4 \log (\nu/{\rm GHz}).
\end{equation}
We considered the scattering effected on the polarized emission, and
found that scattering can indeed flatten the PA curves  \cite{lh04}. The
simulations and the following observations by Camilo et al.  \cite{crj+08}
have confirmed such an effect.

\vspace{-2mm}\subsection{Magnetic field structure of our Galaxy}

For a pulsar at distance $D$ (in pc), the rotation measure (RM, in
radians~m$^{-2}$) is given by
\begin{equation}
{\rm RM} = 0.810 \int_{0}^{D} n_e {\bf B} \cdot d{\bf l},
\end{equation}
where $n_e$ is the electron density in cm$^{-3}$, ${\bf B}$ is the vector
interstellar magnetic field in $\mu$G and $d {\bf l}$ is an elemental vector
along the line of sight from a pulsar toward us (positive RMs correspond to
fields directed toward us) in pc. With the $DM$ we obtain a direct estimate
of the field strength weighted by the local free electron density
% timing is 
\begin{equation}
\langle B_{||} \rangle  = \frac{ \int_{0}^{D} n_e {\bf B} \cdot d{\bf
l}} {\int_{0}^{D} n_e d l}  = 1.232 \frac{RM}{DM},
\label{eq-B}
\end{equation}
where RM and DM are in their usual units of rad m$^{-2}$ and cm$^{-3}$ pc
and $B_{||}$ is in $\mu$G.
Previous analysis of pulsar RM data has often used the model-fitting method
 \cite{hq94,id99}, i.e., to model magnetic field structures in all of the
paths from pulsars to us (observer), and fit the model to the observed RM
data with the electron density model \cite{cl02}. {\it Significant
improvement} can be obtained when both RM and DM data are available for many
pulsars in a given region with similar lines of sight. Measuring the
gradient of RM with distance or DM is the most powerful method of
determining both the direction and magnitude of the large-scale field local
in that particular region of the Galaxy \cite{hml+06}. One can get %
\begin{equation}
\langle B_{||}\rangle_{d1-d0} = 1.232 \frac{\Delta{\rm RM}}{\Delta{\rm DM}},
\label{delta_rm_dm}
\end{equation}
where $\langle B_{||}\rangle_{d1-d0}$ is the mean line-of-sight field
component in $\mu$G for the region between distances $d0$ and $d1$,
$\Delta{\rm RM} = {\rm RM}_{d1} - {\rm RM}_{d0}$ and $\Delta{\rm DM} ={\rm
DM}_{d1} - {\rm DM}_{d0}$. From all available data, we found
that \cite{hml+06} magnetic fields in all inner spiral arms are
counterclockwise when viewed from the North Galactic pole. On the other
hand, at least in the local region and in the inner Galaxy in the fourth
quadrant, there is good evidence that the fields in inter-arm regions are
similarly coherent, but clockwise in orientation. There are at least two or
three reversals in the inner Galaxy, probably occurring near the boundary of
the spiral arms. The magnetic field in the Perseus arm cannot be determined
well. The negative RMs for distant pulsars and extragalactic sources in fact
suggest the inter-arm fields both between the Sagittarius and Perseus arms
and beyond the Perseus arm are predominantly clockwise. See my recent
reviews  \cite{han08,han07} for more details and references therein.

\section{Pulsar timing and gravitational waves}

Pulsars emit pulses with accurate period ($P$).  After time-of-arrival (TOA)
of every pulse is measured (timing observations), one has to correct all
measurements to the barycenter of the Solar system. It is easy to find that
pulsars are slowing down due to radiation. The rate of period change
($\dot{P}$) can be measured. From $P$ and $\dot{P}$, one can estimate the
magnetic field strength on the neutron star surface, $B=3.2\times10^{19}
\sqrt{P \dot{P}}$. 

The timing data usually have some ``noise'', not from measurement
uncertainty but from the stars them-self. More interesting is that from timing
data of young pulsars, there are spectacular changes in the rotation period,
known as ``glitch''. About 100 glitches have been observed from some 30
pulsars  \cite{wmp+00,klg+03,zwm+08}. These glitches provide a penetrating
means for investigation of pulsar interior structure.

Millisecond pulsars are very stable (small $\dot{P}$) and have very tiny
timing-noise. By timing millisecond pulsars, especially pulsars in binary
system provide tools to study gravitational theory. The orbital motion of
pulsars and relativistic effects can be easily measured through the pulse
time delay. The first extra Solar planet was discovered by using timing the
millisecond pulsar PSR B1257+12  \cite{wf92}. The gravitational redshift and
time dilation as well as the Shapiro delay have been detected from a number
of pulsar binary system  \cite{s04}, and the masses of pulsars as well as 
their companions can be measured very accurately.

The most exciting is the discovery of a relativistic binary pulsar, PSR
B1913+16  \cite{ht75} which have been used for examination of the general
relativity. The orbital shrink, because of the gravitational wave radiation,
has been detected from this pulsar -- which reveals a new radiation form in
the nature and was awarded the Nobel Prize in 1993. The newly discovered 
binary pulsar  \cite{lbk+04} can act as the testbed even better \cite{ksm+06}.

\section*{Acknowledgments}
 I am very grateful to Dr. R.T. Gangadhara for inviting me to participate
the ``First Kodai-Trieste Workshop on Plasma Astrophysics'', and for their 
great local hospitality. I wish this lecture-notes can be useful for
students to choose pulsar studies in future. India has many world famous
pulsar astronomers and we wish to make tight connections on pulsar studies
in future, given the fact that the radio telescopes, Indian GMRT and Chinese
FAST, both make pulsar study as primary projects.  My research in China at
present is supported by the National Natural Science Foundation (NNSF) of
China (10521001 and 10773016) and the National Key Basic Research Science
Foundation of China (2007CB815403).

\printindex
\end{document}